\documentclass[showkeywords,amsmath,amssymb]{revtex4}

\usepackage{graphicx}
\usepackage{bm}

\begin{document}

\title{Proof-of-Concept Experiments for Quantum Physics in Space}

\author{Rainer Kaltenbaek, Markus Aspelmeyer, Thomas Jennewein, Caslav Brukner and Anton Zeilinger}
 \affiliation{Institut f\"ur Experimentalphysik, Universit\"at Wien, Boltzmanngasse 5, A-1090 Wien, Austria}
\author{Martin Pfennigbauer and Walter R. Leeb}
 \affiliation{Institut f\"ur Nachrichtentechnik und Hochfrequenztechnik, Technische Universit\"at Wien, Gu\ss haussstra\ss e 25/389, A-1040 Wien, Austria}

\pagestyle{empty}

\begin{abstract}
Quantum physics experiments in space using entangled photons and
satellites are within reach of current technology. We propose a series of
fundamental quantum physics experiments that make advantageous use
of the space infrastructure with specific emphasis on the
satellite-based distribution of entangled photon pairs. The experiments are
feasible already today and will eventually lead to a
Bell-experiment over thousands of kilometers, thus demonstrating
quantum correlations over distances which cannot be achieved by
purely earth-bound experiments.
\end{abstract}



\maketitle


\section{\label{intro}Introduction}

Space provides a unique "lab"-environment for quantum
entanglement: In the case of massive particles, the weak
gravitational interaction enables the expansion of testing
fundamental quantum properties to much more massive particles
than is possible today~\cite{Arndt02}. In the case of photons,
the space environment allows much larger propagation distances
compared to earth-bound free space experiments. This is mainly
due to the lack of atmosphere and due to the fact that space
links do not encounter the problem of obscured line-of-sight by
unwanted objects or due to the curvature of the Earth. Quantum
experiments over long distances are usually based on the
transmission of photons. Earth-based transmission is limited,
however, to some hundred kilometers both for optical
fibers~\cite{Waks02,Gisin02} and for ground-to-ground free-space
links~\cite{Horvath02}. The added value of space will open up new
possibilities for true long-distance experiments based on quantum
entanglement utilizing satellites.

At present, ESA and NASA are hosting five experimental missions
concerned primarily with fundamental physics in space, namely
LISA~\cite{LISA03a}, OPTIS~\cite{OPTIS01a}, GP-B~\cite{GPB03a},
MICROSCOPE~\cite{MICROSCOPE01a} and
STEP~\cite{STEP03a}~\footnote{European Space Agency (ESA);
National Aeronautics and Space Administration (NASA); Laser
Interferometer Space Antenna (LISA); Satellite Based Optical Test
of Special and General Relativity (OPTIS); Gravity Probe-B (GP-B);
MICROSatellite pour l'Observation du Principe d'Equivalence
(MICROSCOPE); Satellite Test of the Equivalence Principle (STEP)}.
We suggest in the following a series of proof-of-concept
demonstrations for quantum physics experiments in space. The
first part of the paper introduces several fundamental tests
concerning both the nature of quantum correlations and the
interplay between quantum physics and relativity. In the second
part, we identify a test of Bell's inequality over astronomical
distances as the first important achievement for
entanglement-based quantum experiments in space. We propose a
series of experiments consisting of three stages, each based on
the other, which will eventually lead to the first
satellite-based demonstration of violating a Bell inequality over
distances that are not feasible with only earth-bound
technology. This will also be of importance for future
applications of novel quantum communication technologies based on
satellites~\cite{Aspelmeyer03a}.

\section{\label{Outview}Fundamental Tests of Quantum Physics in Space}
In the following, we conceive experiments for the demonstration
of fundamental principles of quantum physics, which make
advantageous use of the space infrastructure. Specifically, we
will exploit the possibilities of satellite-distributed quantum
entanglement with photons. Those experiments, although envisioned
to be realizable only as long-term projects, include

\begin{itemize}
\item a Bell experiment using only satellites to demonstrate
quantum correlations over astronomical distances (see Sect.~\ref{sub:Bell1}),
\item a Bell experiment utilizing the freedom of choice of human observers as the necessary random element in choosing the measurement basis(see Sect.~\ref{sub:Bell2}),
\item experiments testing different models considering the
collapse of the wave function as a physical process (see Sect.~\ref{sub:Bell3}),
\item experiments concerning special relativistic and general relativistic
effects on quantum entanglement~(see Sect.~\ref{sub:Bell4}), and
\item Wheeler's delayed choice experiment (see Sect.~\ref{sub:Bell5}).
\end{itemize}

\subsection{\label{section:Bell}Testing Bell-type inequalities}
Classical physics is based on the assumptions of locality and
realism. Reality supposes that results of measurements are
associated to properties that the particles carry prior to and
independent of measurements. Locality supposes that the
measurement results are independent of any action performed at
spacelike separated locations. Local realism imposes certain
constraints on statistical correlations of measurements on
multi-particle systems (Bell inequalities)~\cite{Bell64}. Quantum
mechanics, however, violates the Bell inequalities and is
therefore in contradiction with at least one of the underlying
principles. Up to now, many experiments have been performed
confirming the quantum mechanical predictions (for an overview of
these so-called ''Bell experiments" see for example
\textit{Tittel and Weihs}~\cite{Tittel01a}). To perform such kind
of experiments over long (even astronomic) distances would verify
the validity of quantum physics and the preservation of
entanglement on these scales (see Sec.~\ref{sub:Bell1}). Additionally, a Bell experiment is always the first step towards the experimental realization of entanglement based quantum communication schemes.
Furthermore, a possible decay in the quantum correlations can be
used to test relativistic influences (see Sec.~\ref{sub:Bell4})
and models proposing a physical collapse of the wave function (see
Sec.~\ref{sub:Bell3}).

\subsubsection{\label{sub:Bell1}Bell experiments over long distances}

\begin{figure}

\begin{center}

\includegraphics[width=5cm]{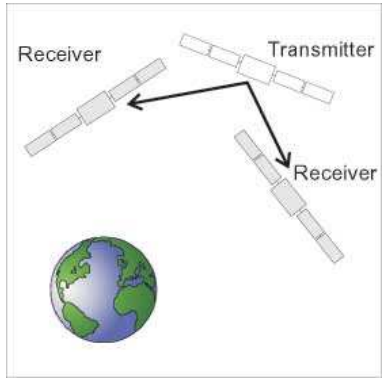}

\end{center}

\caption{In this scheme all three terminals (one transmitter, two
receivers) are placed on satellites. This provides maximal
flexibility for a wide range of experiments over long distances
without any losses due to atmospheric effects. \label{spaceonlyBell}}
\end{figure}

Photons are ideal for propagating over long-distances in vacuum.
The experimental prerequisites to perform Bell experiments are a
source of entangled photons (located in the transmitter terminal) and two
analyzing receiver-terminals, which individually can vary their
measurement basis and store the arrival time of single-photon
detection events with respect to a local time standard.
Specifically, in the case of polarization-entangled photons,
polarization measurements are performed with varying polarizer
settings at each receiver site.
%
To guarantee the independence of the measurements in each of the
receiver-terminals, the measurements have to be space-like
separated. This is more readily accomplished over large distances
between the receiver terminals. In the long run, the optimal solution for a Bell experiment over long distances would be to exclude atmospheric losses by placing both
receiver terminals and the transmitter terminal on
independent satellites (see fig.~\ref{spaceonlyBell}).

This scheme would allow an almost arbitrary variation of the
distances between the different terminals. At the same time,
different relative velocities can be chosen, which is also be
desirable for other experiments proposed in this paper (e.g.~the
experiments utilizing special relativistic effects). The actual
achievable distance is ultimately limited by the size of the
transmitting and receiving telescopes.

\subsubsection{An ultimate Bell experiment\label{sub:Bell2}}
A fully conclusive experiment to test the violation of a Bell
inequality has to obey true randomness in the choice of the
measurement settings: the experimenters' measurement choices have
to be assumed to be uncorrelated with properties of the measured
system prior to measurement (``free will''
criterion)~\cite{Bell85}. Thus far, all experiments utilized
classical or quantum random number automata for the choice of
their measurement. However, in a \textit{completely} deterministic
universe, the free will criterion may not be met, since these
choices of the settings could be conspiratorially correlated with
the properties of the measured system~\footnote{Note, that this
scenario would also make the conceptual distinction between
"locality" and "non-locality" obsolete.}. In order to lead the
determinism-argument completely \textit{ad absurdum}, we suggest
to take the "free-will" criterion literally and involve two human
beings in the Bell experiment, who decide on the choice of the
measurement settings freely and independent from each other. In
this case, a violation of a Bell inequality would imply, for a
deterministic view, that even our free will is conspiratorially
correlated with the properties of the measured system.

\begin{figure}

\begin{center}

\includegraphics[width=10cm]{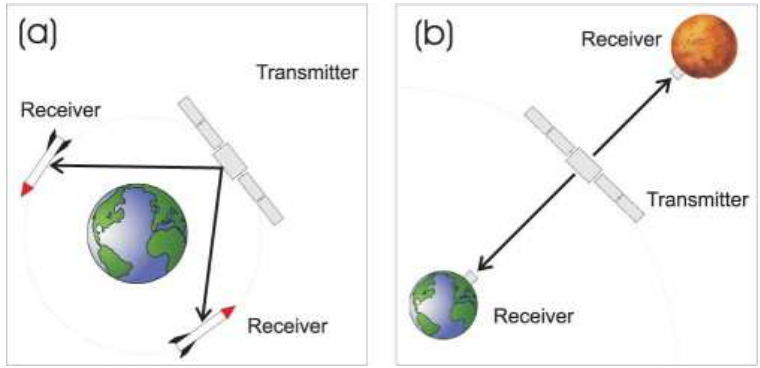}

\end{center}

\caption{(a)Scenario in which two astronauts and the source of entangled photon pairs are separately orbiting Earth. Due to their different propagation velocities there will naturally be periods of time where the necessary distances are reached. (b) Scenario for
a Bell experiment with one astronaut on (or near) Earth, one on
(or on the way to) Mars and the source of the entangled photon pairs on an orbit in between. \label{orbitalBell} }

\end{figure}

To perform this ultimate Bell experiment, two astronauts have to
be placed apart far enough to make sure that their
decisions which measurement to perform are
space-like separated during the experiments and to ensure that they
have sufficient time to make these decisions.
Specifically, if we assume a transmitter terminal emitting
entangled photon pairs mid-way between the astronauts and if we
safely grant each of the astronauts one second of time to make his conscious decision of parameter settings\footnote{We note that it takes of the order of 0.1~s to make a conscious decision.}, the two of them would
have to be separated by at least two light seconds, i.e.~approx.~$600\,000$~km~\footnote{That way the astronaut's free choice is made at a time after the entangled photon pair has been emitted and thus cannot influence the creation of the state.}. To reach the necessary distance between
the two astronauts, it would suffice to place them in opposing
directions at approximately the distance of the moon. One possible
scenario is as follows: the two astronauts and the source are all placed in orbits around Earth such that during some periods of time the distances necessary to perform the  experiment are reached (see Fig.~\ref{orbitalBell}a). Alternatively, it is in principle possible to send only one astronaut to space while the second experimenter stays on Earth. The advantage of a completely space-based scenario is of course the absence of atmospheric losses.

A different scenario could be combined with a future Mars mission. One experimenter accompanies the mission while the second one stays on Earth (or on-board the ISS to exclude atmospheric influences) and the source of entangled photon pairs is sent to an orbit between the orbits of Earth and Mars. As soon as the necessary distances are reached the experiment can be performed (see Fig.~\ref{orbitalBell}b).
\subsubsection{Experiments testing the physical collapse of the wave function\label{sub:Bell3}}

In a quantum measurement, we find the system to be in one of the
eigenstates of the observable defined by the measurement
apparatus. A specific example is the measurement on a wave
packet. Such a wave packet is more or less well-localized, but we
can always perform a position measurement on a wave packet which
is better localized than the dimension of the packet itself. This
so-called ''collapse of  the wave function" is a change in the quantum state of a system which
is sometimes viewed as a real physical process.
%
%
%
Although we do not share the view that the
collapse of the wave function~\footnote{The wave function is a
purely mathematical description of the knowledge about the
system. When the state of a quantum system has a non-zero value at
some position in space at some particular time it does not mean
that the system is physically present at that point, but only that
our knowledge (or lack of knowledge) of the system allows for the
possibility of being present at that point at that instant.} is a physical process, we may still ask with which velocity such a collapse would propagate~\cite{Scarani00}. Experimental tests might exploit the fact that, assuming the
collapse takes place in a preferred reference frame, the
observation of quantum correlations in a moving reference frame
allows to give a lower bound on the speed~\cite{Scarani00}. Present experiments give lower
bounds for the velocity of a potential physical collapse up to
$10^7$ times the speed of light~\cite{Zbinden01a}. Bringing such
experiments to space could drastically expand the testable scale,
primarily due to the large distances involved and the high speeds
of the satellites.

\subsection{Tests of special and general relativistic effects on quantum
entanglement\label{sub:Bell4}}

\subsubsection{\label{Effects}Experiments involving special relativistic effects}

Due to the potentially high velocities and large distances in
space experiments, it might be of interest to consider possible
relativistic effects on entanglement, although it is obvious that
these effects will not be dominating. A recent overview on many of
these effects has been given by \textit{Peres and
Terno}~\cite{Peres02a}.

Recent research shows that the entanglement of
polarization-entangled photon pairs depends in general on the
observers' reference frame~\cite{Bergou03a}, in other words,
polarization entanglement alone is not a Lorentz-invariant
scalar. Yet, the overall entanglement in the full Hilbert space
of the two photons is preserved under Lorentz transformation,
which means that entanglement is effectively transferred between
the degrees of freedom polarization and
momentum~\cite{Milburn02a,Gingrich02a,Li02a}. Similar effects can
also be observed for massive particles between spin and momentum.
Note that, in a standard lab experiment, such transformations
would require the use of optical elements such as polarizing
beamsplitters~\cite{Zukowski88}.

%
%
To test the behaviour of entanglement under Lorentz
transformations, scenarios have to be found in which the relative
velocities between observers and a transmitter terminal carrying
the entangled source is high enough to allow for the measurement
of special relativistic effects. To arrive at high relative
velocities, the \textit{space--to--space} scenario is again the
most flexible one, also since all the other Bell experiments can
easily be performed using the same resources.


\subsubsection{\label{ART}Experiments involving general relativistic effects}
When sharing entanglement over distances comparable to or greater than the distance Sun~-~Earth, one
has to consider the possibility of gravitational influences on
entanglement.

Polarization- and spin-entanglement leads to correlations between the outcomes of
polarization (or spin) measurements on both of the particles.
Such measurements however, can only have an unambiguous
operational meaning if directions like "horizontal" or "vertical"
("up" or "down") on each side are well defined. Many experimental schemes for quantum communication (e.g.~quantum key distribution) require a common reference frame between the observers \footnote{A counter-example are Bell-type experiments, which do not need a common reference frame.}. For two particles
moving apart, the initially joint reference frames, which yield
perfect correlations will be parallel-transported along their
individual trajectories. In general, however, quantum particles
need not be associated with a unique trajectory. Therefore, one
has to take into account all paths a particle can possibly take
and sum up the effects of gravity on the particle along these
ways weighted by their probability-amplitudes. For each path, the
reference frame yielding perfect correlations will be slightly
different. Recently it was suggested, that this can lead to a
decrease in the correlations between two
particles~\cite{Mensky00a,Terashima03}. Bell-experiments over
sufficiently large distances might be able to demonstrate such
effects although up to now the work on this field is purely conceptual and no theoretical predictions have been made as to quantify the expected decrease in quantum correlations in actual experiments.

\subsubsection{Entanglement-Enhanced
Interferometry\label{subsec:entint}} Quantum entanglement allows
to effectively increase the phase-sensitivity $\Delta\Phi$ of
interferometers. By preparing specific photon-number entangled states in the
arms of an interferometer $\Delta\Phi$ can be improved quadratically
from $1/\sqrt{N}$ to $1/N$, where N is the number of photons in
the input state entering the
interferometer~\cite{Caves80a,Jacobson95}. Dowling~\cite{Dowling98a} derived a general formalism valid for fermions and bosons and provided estimates for the performance of optical, atom-beam, and
atom-laser interferometers~\footnote{Note, that a direct comparison between
atomic and optical interferometers has to take into account the
momentum, the flux and the roundtrip time (time needed for a particle to pass through the interferometer) of the particles involved. While photons have clear disadvantages concerning their
momentum, atoms suffer from low flux and long roundtrip times.}. For example, an optical entanglement-enhanced interferometer might be up to
$10^8$ times more sensitive than a regular interferometer for the same number of photons passing through the interferometer~\footnote{With today's state of the art technology, however, the possible flux achievable for single photon interferometers is incomparably higher than for entanglement-enhanced interferometers.}.

\paragraph{Testing the Lense-Thirring effect using
entanglement\label{sub:Bell6}} Though general relativity is in
many ways an established theory there are still some of its
central predictions that need accurate testing like the
Lense-Thirring effect~\cite{Lense18}. This effect is due to the
dragging of inertial frames in the vicinity of rotating
gravitating bodies like Earth, which induces an anisotropy in the
surrounding space-time~\cite{Misner73a,Ciufolini95a}. First
experimental evidence has already been collected by the LAGEOS
and LAGEOS II satellites~\cite{Ciufolini98a}. In general, this
anisotropy is experimentally accessible via the Sagnac effect, by
which a preferred direction in an interferometer creates a phase
shift between interfering modes (see e.g.
\textit{Stedman}~\cite{Stedman97a}). Specifically, a weak
gravitational field results in a phase-shift
\begin{equation}
\Delta \Psi = -\frac{4 \pi}{\lambda}\oint dx^{i} h_{0i}, \label{sagnac1}
\end{equation}
where $\lambda$ is the mean wavelength in the absence of rotation.
$h_{\mu \nu}$ describes the deviation of the metric tensor from
the Minkowski-metric ($\eta_{\mu \nu}$) and thus a small
deviation from flat space-time: $g_{\mu \nu} = \eta_{\mu \nu} +
h_{\mu \nu}$. This is an important objective of the
HYPER-mission, which will be using "hyper"-cold-atom
interferometers with increased experimental resolution to obtain
more accurate measurements of the Lense-Thirring effect. An
additional increase is to be expected by the use of
entanglement-enhanced interferometry (see
Sec.~\ref{subsec:entint}). A general requirement for such
experiments would be an optical interferometer in space. Since
large optical distances should be covered, one can ideally
imagine a flotilla configuration of satellites, where optical
path lengths can be stabilized down to sub-wavelength scales.
Such flotilla configurations are currently being investigated by
ESA and NASA as high-resolution telescopes based on nulling
interferometry~\cite{Darwin}.


\paragraph{Testing G\"odel's cosmological model\label{sub:Bell7}}

In 1949 G\"odel suggested an alternative solution for Einstein's
field equations of gravitation by assuming a net rotation of the
universe as a whole~\cite{Godel49}. It was recently pointed out by
\textit{Delgado }\textit{et al.}~\cite{Suessmann02a}, how an
entanglement-enhanced interferometric resolution might lead to the
possibility to experimentally test for G\"odel's cosmological
model. Since rotating masses are involved, the experimental
scheme is naturally equivalent to the one used for the test of
the Lense-Thirring effect (see above). Taking into account the
overall mass density distribution~(Delgado et al. assume
$\rho=2\cdot10^{-31}g~cm^{-3}$) one can predict the rotation-rate
for the universe to be $\Omega_U \cong 4 \times 10^{-19}
rad~s^{-1}$, which is still small compared to the rotation rate
corresponding to, for example, the Earth's Lense-Thirring effect
$\Omega_{TL} \cong 10^{-14} rad~s^{-1}$~\cite{Suessmann02a}.
Currently, without using entanglement, the best achievable
accuracy is approx. $10^{-16} rad~s^{-1}$ calculated for the
HYPER-interferometer in the course of one year. This means, an
experimental run to test for G\"odel's model would take some 1000
years to be accomplished. However, taking into account the
possibility of entanglement-enhanced interferometry would
possibly allow for much shorter, eventually experimentally
feasible integration times.

\subsection{Wheeler's Delayed Choice experiment\label{sub:Bell5}}

\begin{figure}

\begin{center}

\includegraphics[width=6cm]{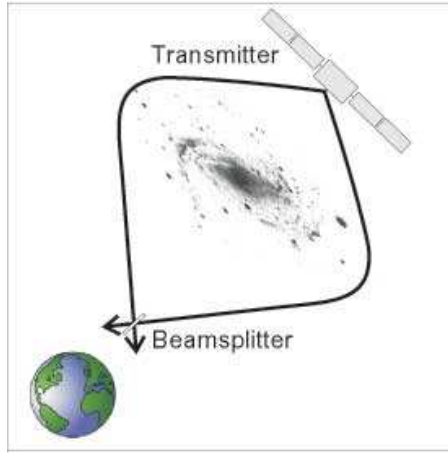}

\end{center}

\caption{Wheeler's delayed choice experiment: the photons coming from a far source are deflected by a
strong gravitational lens and brought together again to be either
detected immediately or after overlapping on a beam-splitter.
\label{wheelerBild}}

\end{figure}

Delayed choice experiments show in an impressive way how physical
realism, namely that particles have definite paths (position and
momentum are equally well defined), leads to contradictions with
either quantum physics or locality (see e.g. \textit{Hellmuth et
al.}~\cite{Hellmuth87a}). They are generally based on the fact,
that, depending on the experimental setup, a quantum system can
either exist in a superposition of two orthogonal states or is
"localized" in one of the two states. However, the decision on
the actual setup can be delayed way beyond the time, when the
system enters the experimental apparatus. This contradicts
the realist notion that the properties of a physical system are
predetermined during the \textit{whole }time of the experiment. Taking physical
intuition to the extreme, John Wheeler proposed an experiment, in
which the experimenter's intervention might be delayed even by
some millions of years~\cite{Wheeler78a}: Suppose there is a light
source at an astronomical distance, emitting single photons in our
direction. All these photons have to pass a gravitational lens
(e.g. a galaxy or a black hole) in a way that the possible paths
(to the right and to the left of the massive object) cross in the
vicinity of Earth (see fig.~\ref{wheelerBild}). One may now ask,
how the photons will behave depending on the experimenter placing
a beamsplitter at the intersection or not. Quantum physics
predicts that the presence or absence of interference only
depends on the actual positioning of the beamsplitter as decided by
the experimenter, although in a pure realist's particle-picture
the particle had to choose its way maybe millions of years ago. In other words, the realist seems to be "deciding what the photon shall have done after it has already done it!"~\cite{Miller83a}.

For an experimental realization it would not be feasible to bring
a single-photon light source to the other end of a galaxy to test
these assumptions. However, there have indeed been discovered
objects we see twice as their light passes on two sides of a
gravitational lens \cite{Schneider92a,Walsh79a}. In principle, all one would
have to do is placing a beamsplitter in the intersection of the two
paths and try to observe interference. Practically, the two possible ways of the photons will slightly differ, i.e.~their optical paths will differ by a few lightyears. Therefore one would have to introduce a delay-loop in one of the arms to allow for interference. Obviously, a loop delaying a photon for five years might be hard to realize. 


\section{\label{POPexperiments}Feasible proof-of-principle experiments}
After having discussed experiments for fundamental tests of
quantum physics in space, we will now provide a roadmap
towards first feasible demonstrations of the underlying concepts.
This will eventually lead to an experiment in which quantum
entanglement is distributed between two ground stations via a
satellite link.

\begin{figure}
\begin{center}
\includegraphics[width=13cm]{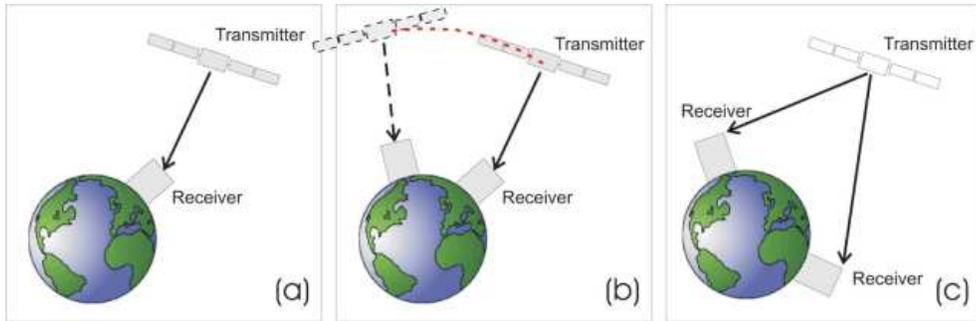}
\caption{\label{fig:overview}Scenarios for feasible
proof-of-concept experiments utilizing entangled photon pairs and
satellites.}
\end{center}
\end{figure}

It has been argued recently that state-of-the-art technology can
already be used to exchange
single-photons~\cite{Nordholt02,Rarity02} or even entangled photon
pairs~\cite{Aspelmeyer03a} via optical free-space links between
satellites and/or ground stations thus allowing novel quantum
communication protocols such as quantum cryptography in a space setting. The next
step is to perform proof-of-principle experiments
actually testing these findings using satellite-to-ground links.
Both, the distribution of ''single-photon" faint-laser pulses and
the distribution of entanglement via terrestrial optical
free-space links has already been verified experimentally over
considerable distances~\cite{Hughes1,Kurtsiefer1,Aspelmeyer03c}.
With this in mind we propose three space experiments which will
eventually lead to a long-distance Bell experiment. The
realization of these experimental schemes is of a modular nature
where the source of entanglement is placed within a space-borne
transmitter terminal and the measuring units are placed in
independent receiver terminals at ground stations. In a first
experiment, single photons are sent from the space-borne
transmitter terminal to a ground station to demonstrate
single-photon quantum cryptography (see
fig.~\ref{fig:overview}a). In a second experiment, this scheme is
used to demonstrate secure quantum key distribution between two
\textit{arbitrary }ground stations by independently establishing
a key at each pass of the satellite (see
fig.~\ref{fig:overview}b). In a third experiment, entangled
photons are distributed to two ground stations simultaneously
allowing the violation of a Bell's inequality between independent
ground stations separated by more than 1600 kilometers (see
fig.~\ref{fig:overview}c).

\subsection{Technological baseline and preliminary design}
All proposed experiments are based on the preparation and
manipulation of photonic entanglement. We will focus here on
polarization-entangled photon pairs, since this is most suited for
the propagation through the non-birefringent
atmosphere~\cite{Aspelmeyer03a}.


\subsubsection{Transmitter}
\begin{figure}
\begin{center}
\includegraphics[width=13cm]{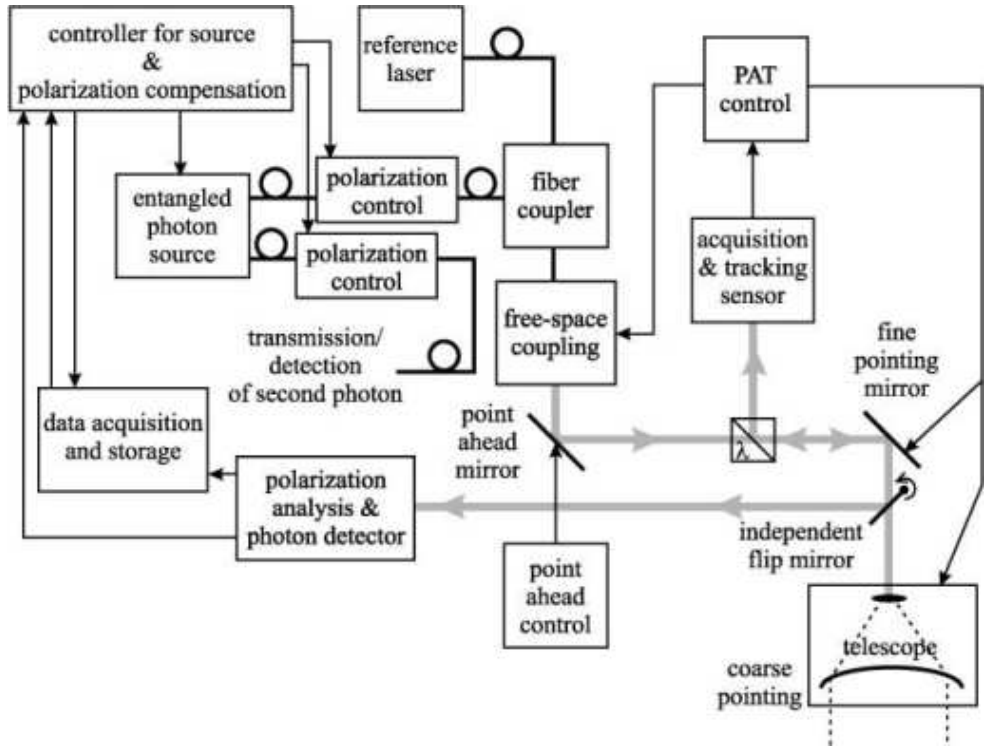}
\caption{\label{fig:transmitter_block}Block diagram of the
transmitter terminal.}
\end{center}
\end{figure}

The transmitter terminal comprises the entangled photon source,
modules for polarization-sensitive manipulation and detection of
single photons, and a telescope combined with an optical pointing,
acquisition and tracking~(PAT) system ~\footnote{PAT systems are
used to establish and maintain the line-of-sight connection in
optical free-space links when the terminals are moving relatively
to each other.} to establish the downlink(see Fig.~\ref{fig:transmitter_block}). The photons of each entangled pair
are coupled into optical fibers, which allow polarization control
via (piezomechanical) bending of the fibers. Coupling to the
classical optical head is then achieved via a fiber coupler.
Depending on the stage of the experiment, the two photons of the
entangled pair are either both transmitted through separate
telescopes or only one is transmitted while the other one is
immediately detected. The entangled photon source subsystem
comprises additional laser diodes for alignment of the optical
fibers. The reference laser of the PAT subsystem is linearly
polarized and optionally pulsed to provide both an orientational
and a timing reference frame between transmitter and receiver
site. A point ahead angle unit (PAA) provides the required
non-parallelity between the optical axes of the downlink and the
uplink.

We distinguish the following modes of operation of the transmitter
terminal: (i)~\textit{Standby mode}~(A closed optical loop for internal
alignment purposes is operational when no downlink is
established), (ii)\textit{PAT mode}~(When a link is available in
principle, the PAT sequence is initiated.), and
(iii)~\textit{Quantum communication mode}~(When the downlink is
available, the entangled photon source is operating with the
alignment laser diodes being inactive.).

\subsubsection*{Receiver}
\begin{figure}
\begin{center}
\includegraphics[width=13cm]{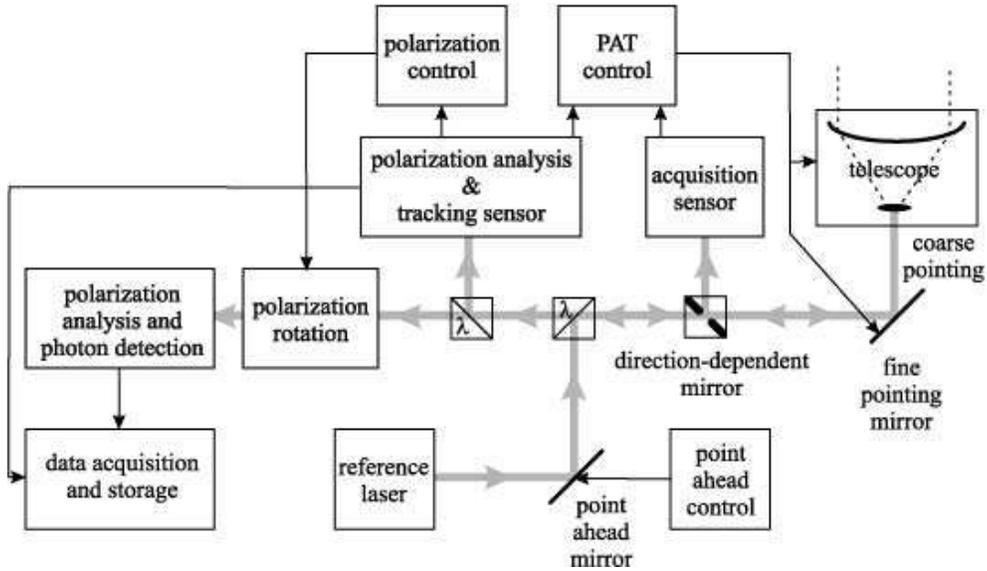}
\caption{\label{fig:receiver_block}Block diagram of the receiver
terminal.}
\end{center}
\end{figure}

The receiver terminal comprises a single-photon analysis and
detection subsystem (analogous to the unit used in the transmitter
terminal) and an optical subsystem consisting of a telescope and a
PAT unit (see Fig.~\ref{fig:receiver_block}). Another
polarization analysis subsystem monitors the polarization of the
transmitter reference laser, which is used to compensate for any
orientational misalignment between transmitter and receiver
polarization. The polarization analysis based on the reference
laser signal does not require the use of single-photon detectors
due to the high intensity of the beam to be analyzed. The signal
from this analysis is used to properly orient the polarization
in the single-photon beam path. The reference laser of the
receiver station(s) is operating at a  wavelength differing from
the transmitter reference laser in order to keep cross talk
sufficiently low.

The modes of operation of the receiver terminal are as
follows~\footnote{For the sake of simplicity, we consider only one
of the photons of the entangled pair.}: (i)~\textit{Standby
mode}~(When the transmitter terminal is in \textit{standby mode},
the receiver terminal is not operating.), (ii)~\textit{PAT
mode}~(When a link is available in principle, the PAT sequence is
initiated.), and (iii)~\textit{Quantum communication mode}~(When
the downlink is established, the beam received from the reference
laser of the transmitter terminal is separated by a dichroic
mirror from the single-photon data beam. During the availability
of the downlink, data is acquired and stored locally with respect
to an accurate local time standard.).


\subsection{Experiment 1: Single downlink\label{sec:Exp1}}

For the first experiment, the transmitter terminal is suggested
to be placed onto a low-Earth-orbiting (LEO) platform, while one
receiver terminal is installed at an optical ground station.
Positioning of the transmitter terminal is critical insofar as it
requires nadir pointing of the module. One possible platform
which fulfills this requirement is one of the external payload
facility stations at ESA's Columbus module hosted by the
International Space Station~(ISS)~\cite{esa01_1}. The receiver
module is located at an optical ground station such as ESA's station (OGS) at Tenerife which
will have to be adapted to properly interface with the receiver.

\subsubsection{Availability of ISS--to--ground link\label{subsec:avail_exp1}}

\begin{figure}[ht]
      \begin{center}
         \includegraphics[width=0.5\linewidth]{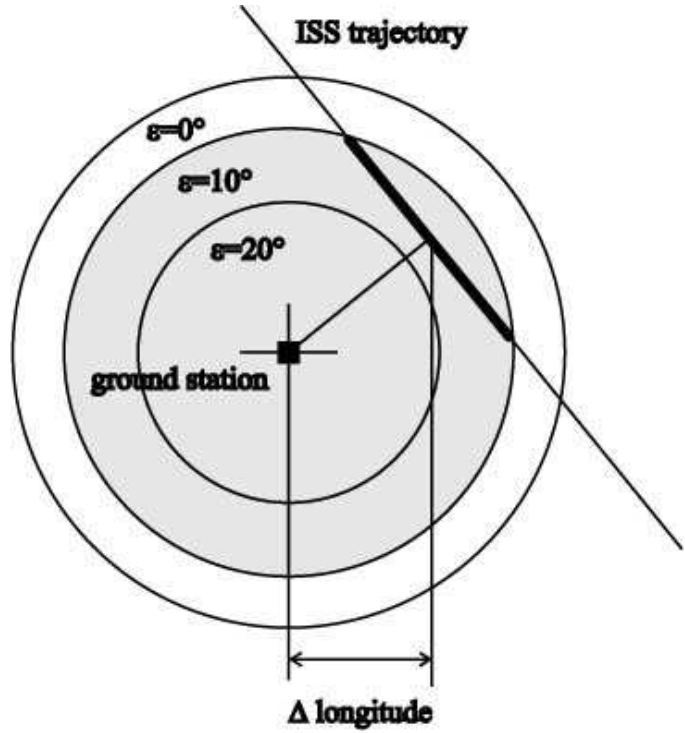}
         \caption{Trajectory of ISS over the field-of-view of a ground station.
         The possible duration of communication depends on the minimum elevation
         angle $\varepsilon$ and the longitudinal shift ($\Delta$ longitude) between
         the ground station and the space station in its zenith.}
         \label{fig:ISS_1station}
      \end{center}
   \end{figure}

\begin{figure}[ht]
      \begin{center}
         \includegraphics[width=0.8\linewidth]{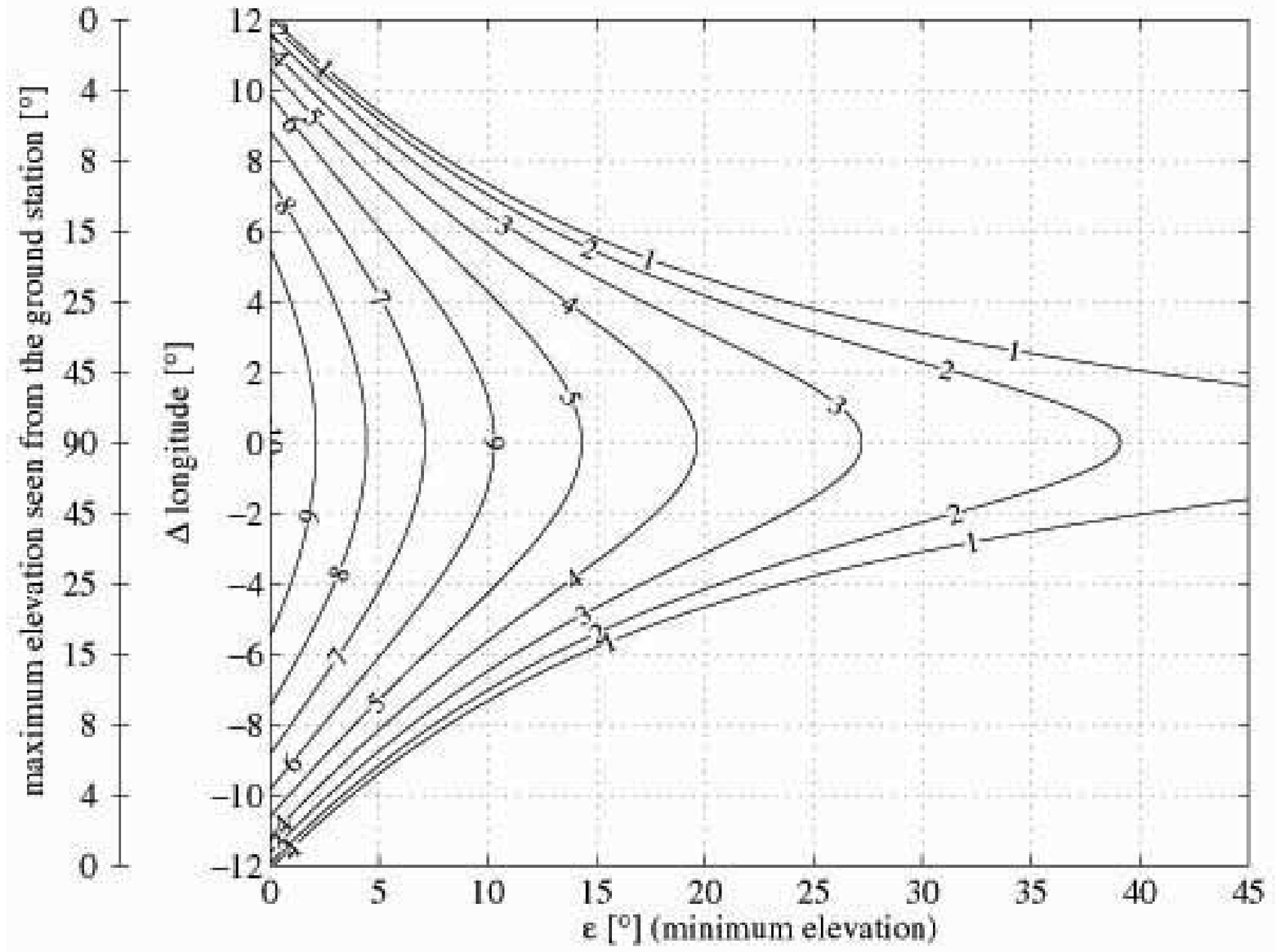}
         \caption{Maximum duration of communication between ISS and the OGS (given
         in minutes by the numbers inserted along the lines) as a function of the
         elevation angle and the difference in longitude of the ground station and the
         satellite when in zenith.}
         \label{fig:ISS_1station_calc}
      \end{center}
   \end{figure}

The ISS orbits at an altitude of around $400~km$, the inclination
angle is $51^\circ$ and one orbit lasts for 92 minutes. The
possible duration of a communication link depends on the height
above sea level, the geographical latitude and altitude of the
ground station and the minimum elevation angle~\footnote{The
elevation angle for which communication is possible
   may be restricted by atmospheric conditions or limited line of sight near
   groun.}. Figure \ref{fig:ISS_1station} shows schematically the trajectory of
   the ISS over a ground station. We have highlighted the part where a link is
   possible, i.e.~for elevation angles exceeding a certain minimum value.
   The link duration for a certain ground station does not only depend on the
   elevation angle but also on the longitudinal shift ($\Delta$ longitude) of the
   individual pass-over. In Fig.~\ref{fig:ISS_1station_calc} corresponding numerical
   values for the OGS can be read off. The results for other
   ground stations mentioned in Section~\ref{subsec:link_exp3} do not differ
   significantly. Orbits with a value for $\Delta$~longitude $< 25^\circ$ will yield
   useful link durations.
   This will be experienced for some 14\% of all the orbits. With 15 orbits per
   day this results in two useful links within 24 hours.

\subsubsection{Description of experiment\label{subsec:description_exp1}}

In the first stage, only one of the photons of an entangled pair
is transmitted to a ground station while the other photon is
directed to the polarization-analysis subsystem, where it is
detected with respect to one of four polarization states (out of
two non-orthogonal bases, i.e.~for example
\{0$\,^{\circ}$,90$\,^{\circ}$\} and
\{45$\,^{\circ}$,135$\,^{\circ}$\}). Its detection also serves as
a trigger event, indicating that a single photon is emitted along
the downlink. Polarization and detection time (with respect to a
local time standard) are locally stored.

At the receiver, two signals generated by the transmitter have to
be processed: the single photons representing the quantum
communication and the reference laser beam. They are spatially
separated by making use of their different wavelengths (single
photons: $\lambda \approx 800~nm$, reference laser: $\lambda
\approx 950~nm$). Since the reference laser is linearly polarized,
a polarization analysis of the transmitted beam could be used to
find the relative rotational orientation between transmitter and
receiver terminal (Note that it is useful, though not always
mandatory, to establish a common reference frame for the
polarizers onboard the transmitter and for those of the receiver.
This can easily be done by calibrating all linear optical
elements with respect to a laboratory reference frame.). The
corresponding feedback signal is used to control the state of
polarization (SOP) in the single photon beam to compensate for
orientational misalignment. The single photons in the quantum
communication channel are then randomly detected with respect to
one of two polarization bases, which are chosen equivalent to the
bases used on-board the transmitter terminal
(e.g.~\{0$\,^{\circ}$,90$\,^{\circ}$\} and
\{45$\,^{\circ}$,135$\,^{\circ}$\}). Polarization and detection
time (with respect to a local time standard) are locally stored.

With a presumed pair generation rate of $500~000$ per second and
an estimated loss of $6.5~dB$ for the local detection of
qubits~\footnote{This is a result of optical losses and finite
detection efficiency.} one can expect a count rate of approx.
$112~000~s^{-1}$ single photons within the transmitter terminal
itself. With the total attenuation of $25dB+6.5dB=31.5dB$ for the
downlink~\cite{Aspelmeyer03a}, we arrive at a count rate of
approx. $350 s^{-1}$ caused by the qubits at the receiver
terminal. As outlined in an earlier paper~\cite{Aspelmeyer03a},
we assume a total background count rate of approx.$1000s^{-1}$ for night-time operation,
which leads to 1350 counts~$s^{-1}$ for the entire detection
process. The final signal rate (defined by the number of joint
detection events at the transmitter and receiver terminal) is
then expected to be approx.80$s^{-1}$ (i.e. the pair generation
rate reduced by the total link attenuation of
$31.5dB+6.5dB=38dB$). For a link duration of 300 seconds this
accumulates to a net qubit transmission of 2400 qubits. One can
expect erroneous coincident detection events on the order of $2
s^{-1}$, which yields a bit error of approx.~2.5\%.

After the experiment, the local data is corrected for varying
signal propagation times and for varying local time standards (the
latter should be negligible when atomic clocks are being used,
while the signal propagation time could be monitored during the
experiment by periodically pulsing the transmitter reference
laser. This would allow to take into account a varying
transmitter-receiver distance along the orbit.). After this data
correction the data sets will be compared with each other to
obtain first information about the link quality such as
efficiency and atmospheric effects. Eventually, a BB84 quantum
key distribution protocol~\cite{Bennett84} can be established by
openly comparing certain subsets of the locally stored data sets.
Then the net key bit rate and the quantum bit error rate (QBER)
of the quantum key distribution protocol will be evaluated. Given
the above approximations, a raw key generation rate of 1.2~kbit
per link duration of approx.~5 minutes might be expected, since
in only half of the cases the joint measurements on the photon
pair will have been performed along the same basis at the
receiver and the transmitter.

\subsection{Experiment 2: Two independent single-photon downlinks\label{sec:Exp2}}

Experiment 2 will establish a quantum key exchange between two
independent ground stations over distances not feasible with
Earth-bound technology. No modification of the space module is
required but a second (independent) ground station has to be
equipped with an additional receiver module. This second ground
station can be located at any arbitrary global position which
allows optical contact with the ISS.

\subsubsection{Description of experiment}
In order to achieve a key
exchange between separate ground stations via ISS, each of the two
ground stations will independently establish a quantum key with
the space-based transmitter terminal as is described for
Experiment~1 (see Section~\ref{subsec:description_exp1}). Since
the space platform has access to both keys, it can send a logical
combination of the keys (e.g.~logically connected by XOR) via classical communication
channels publicly (i.e. not secured) to either of the ground
stations, where the key of the other ground station can be
generated. In principle, a key exchange can thus be achieved
between arbitrary ground stations. However, in this scenario the
security requirement upon the transmitter terminal is as high as
for the ground station. This requirement could be overcome if one distributes
entangled qubits directly to
different ground stations (see Experiment~3,
Section~\ref{sec:Exp3}). In this case, the satellite does not
obtain any knowledge about the distributed key.

With respect to qubit transmission and key exchange rates the
same estimates apply as above for Experiment 1 (see Sect.~\ref{subsec:description_exp1}).

\subsection{Experiment 3: Simultaneous entangled-photon downlink\label{sec:Exp3}}

A highly desirable prerequisite for Experiment~3 is a successful
completion of all stages of Experiments~1 and 2. This includes the
successful establishment and characterization of a downlink
quantum channel as well as the realization of the BB84 quantum
cryptography protocol. In Experiment~3, timing and orientational
synchronization has to be established simultaneously between two
ground stations and the ISS. This represents another degree of
complexity compared to Experiment~2, where the two ground stations
have been addressed independently. However, Experiment~3 will
allow a test of Bell's inequality over distances only achievable
with space technology, and, additionally, a demonstration of
quantum key distribution based on entanglement~\cite{Ekert91a} which 
relaxes the security requirements for the satellite significantly as 
in this case the satellite would not hold any information about the generated key.

The locations of the two ground stations have to be chosen in such
a way that a simultaneous link between ISS and the ground stations
can be established (see Section~\ref{subsec:link_exp3}). In order
to demonstrate features unique for this experiment, the distance
between the two ground stations should be chosen sufficiently
large, i.e. beyond distances which can be bridged with optical
fiber technology (which is limited to a few $100~km$). At the same
time, a modification of the transmitter terminal is required. It
now has to include two separate telescopes together with two
independent PAT subsystems. This design could already be applied
in Experiments~1 and 2 by using a flip mirror in one of the
telescopes' input ports to direct the photon beam to the analyzer
subsystems. With such a transmitter design, no further
modification would be necessary when proceeding from Experiments
1 and 2 to Experiment 3.

\subsubsection{Simultaneous link between ISS and two ground stations\label{subsec:link_exp3}}

\begin{figure}[ht]
      \begin{center}
         \includegraphics[width=0.5\linewidth]{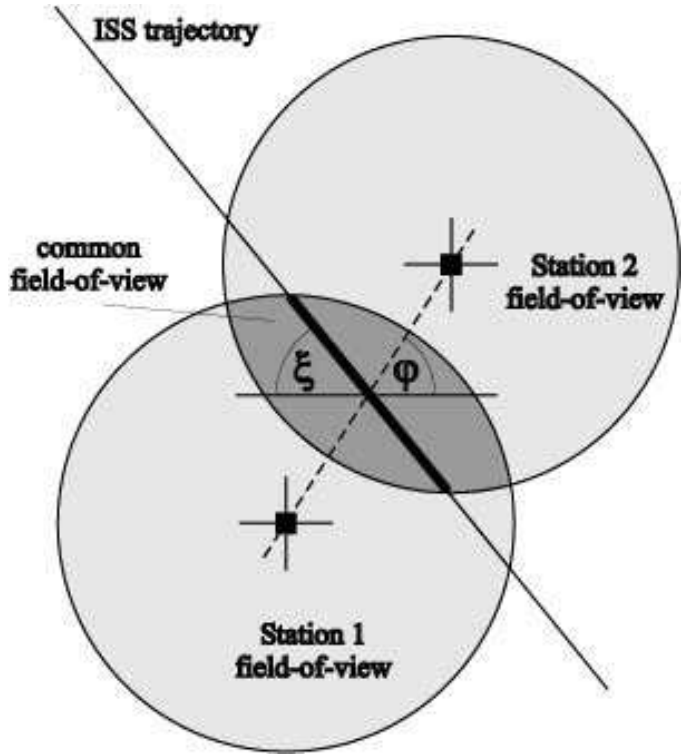}
         \caption{Trajectory of ISS over the field-of-view of two ground stations
         (the angles $\varphi$ and $\xi$ are measured with respect to a line parallel
         to the equator).}
         \label{fig:ISS_2stations}
      \end{center}
   \end{figure}

The scenario illustrating
possible simultaneous links between the ISS and two
   ground stations is presented in Fig.~\ref{fig:ISS_2stations}.
   The link duration now depends on the distance between the
   stations, the angles $\varphi$ and $\xi$ (resulting from the geographical
   position of the two stations), and the minimum acceptable elevation angle.
   Table \ref{tab:dist_stations} gives the corresponding values for
   four representative scenarios.

   \begin{table}[ht]
      \begin{center}
         \begin{tabular}{|c|r|r|r|}\hline
                                                       & distance          & $\varphi$ \hspace{4mm} & $\xi$ \hspace{3mm} \\\hline\hline
            Tenerife   $\leftrightarrow$ Calar Alto    & $1638$~km & $40.9$~°       & $28.0$~°   \\
            Tenerife   $\leftrightarrow$ Matera        & $3309$~km & $33.3$~°       & $25.7$~°   \\
            Calar Alto $\leftrightarrow$ Matera        & $1698$~km & $19.0$~°       & $19.4$~°   \\
            Calar Alto $\leftrightarrow$ Sierra Nevada & $76$~km   & $166.8$~°      & $22.0$~°   \\\hline
         \end{tabular}
      \end{center}
      \caption{Distances between ground stations and angles $\varphi$ and $\xi$ as shown in Fig.~\ref{fig:ISS_2stations}.}
      \label{tab:dist_stations}
   \end{table}

\begin{figure}[ht]
      \begin{center}
         \includegraphics[width=0.7\linewidth]{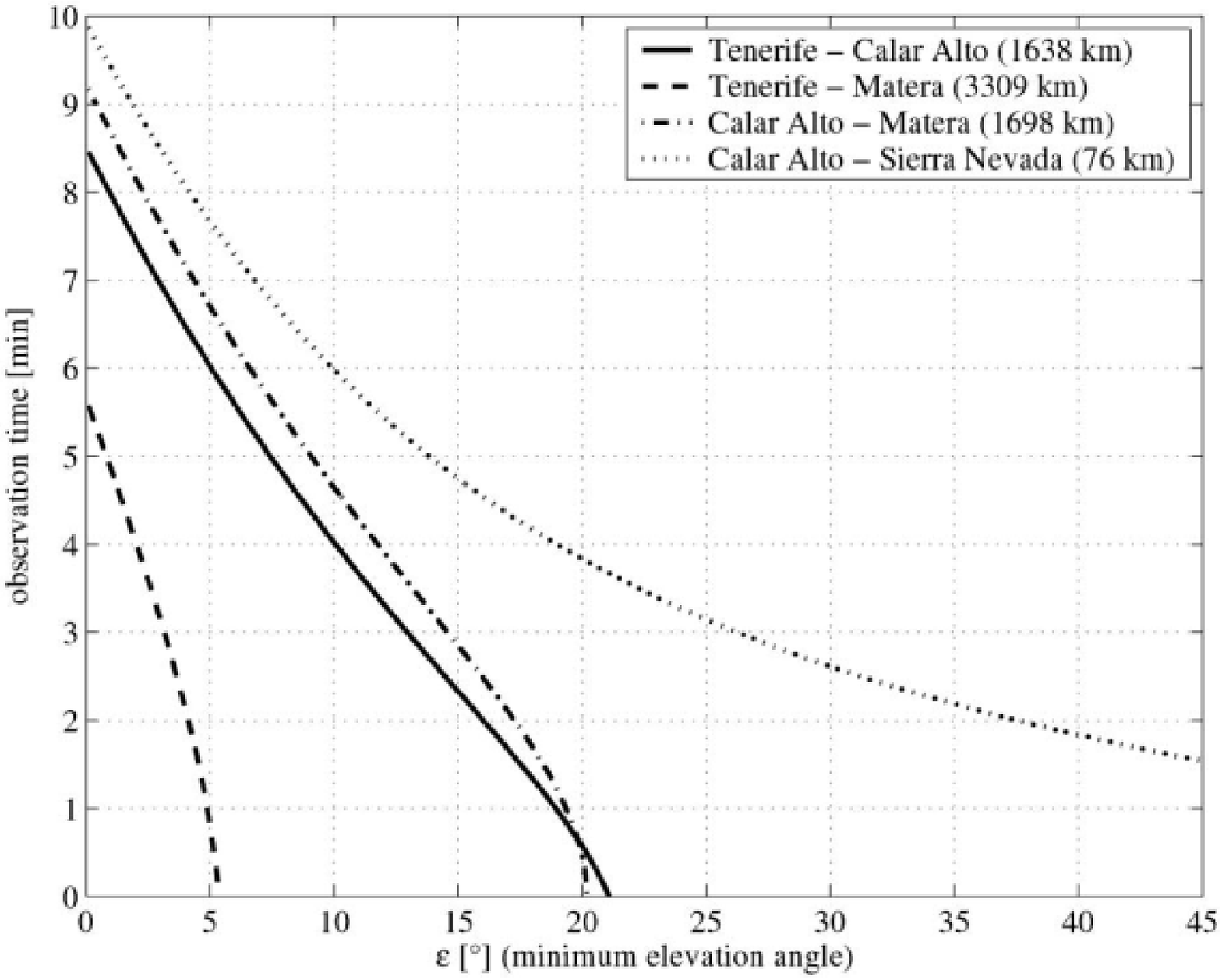}
         \caption{Maximum duration of simultaneous communication between ISS and
         two ground stations as a function of the elevation angle. (The distance cited
         in the insert is that between the stations.)}
         \label{fig:ISS_2stations_calc}
      \end{center}
   \end{figure}

   In Fig.~\ref{fig:ISS_2stations_calc} we present link durations for the optimum case,
   where the ISS passes both ground stations in the 
   symmetric way indicated in Fig.~\ref{fig:ISS_2stations}.
   For each scenario a maximum range of longitude
   for which a link can be established
   results from the geographical position of the 
   stations involved. It is some $25^\circ$ for the link with
   Calar Alto and Tenerife or Matera, some $10^\circ$ for
   Tenerife and Matera. The two stations in Calar Alto and Sierra
   Nevada are so close to each other that the longitudinal range
   is $36^\circ$. The corresponding link rates are
   tow per day for Calar Alto and Tenerife or Matera, one for Tenerife and
   Matera and three per day for Calar Alto and Sierra Nevada.

\subsubsection{Description of experiment}\label{subsec:description_exp3}

Both photons of the
entangled pair are now transmitted simultaneously to two separate ground
stations via independent telescopes. Therefore, at the
transmitter, no analysis and detection of single photons takes
place during quantum communication.

The situation at the receiver stations is equivalent to
Experiment~1 (see Section~\ref{subsec:description_exp1}). At each
of them, the single-photon quantum communication signal
is separated from the reference laser beam, where the
latter serves both as a rotational reference with respect to the
transmitter terminal (i.e.~it provides a control signal to actively
achieve proper orientation of polarization by rotating the
polarization in the quantum communication channel) and allows for
signal propagation measurement to correct for varying link
distance along the orbit. The single photons from the quantum
communication channel are then detected with respect to one of two
(non-orthogonal) polarization bases. Polarization and detection
time (with respect to a local time standard) are locally stored.
Before a comparison of the locally acquired data can take place,
the data is corrected for varying signal propagation times and
varying local time standards.

With an estimated total link attenuation of approx. $31.5dB$ (we
assume a loss of $25dB+6.5dB=31.5dB$ for each of the downlinks),
one can, in the ideal case, expect a count rate of approx.~$1350
s^{-1}$ single photon counts in total at each of the receiver
terminals for night-time operation (here, too, this number includes background radiation).
The final coincidence rate (defined by the number of joint
detection events between the two receiver terminals) is then
calculated by attenuating the available $500~000$ counts/s by $63
dB$, yielding 0.25 counts/s. For a link duration of 300 seconds
this accumulates to a net bit transmission of 75 bits. One
can expect erroneous coincident detection events on the order of
2.5 per 100 seconds, which results in a bit error of approx.~10\%.

The first step to be performed will be a test of Bell's
inequality between the two ground stations. Note once more that,
due to the large distance between the two ground stations, such a
test cannot be performed without employing an Earth-orbiting
satellite. In a first set of measurements, the entangled quantum
state is characterized with respect to its polarization
correlations by keeping the analyzer bases at both receiving
stations fixed at \{0$\,^{\circ}$,90$\,^{\circ}$\} and
\{45$\,^{\circ}$,135$\,^{\circ}$\} (except for a varying
compensation of the transmitter rotation). When comparing the
data sets of the two receiver stations, all the joint detection
events measured with analyzer settings with parallel and
45$^{\circ}$ relative orientation (eight probabilities for joint
events in total) provide already certain bounds for the degree of
purity and entanglement of the state.

In a further set of measurements, one of the receiver stations
keeps the orientation of its analyzing bases fixed at
\{0$\,^{\circ}$,90$^{\circ}$\} and
\{45$\,^{\circ}$,135$\,^{\circ}$\}, while at the other station the
analyzing bases are rotated by 22.5$\,^{\circ}$, resulting in the
basis sets \{22.5$^{\circ}$,112.5$^{\circ}$\} and
\{67.5$^{\circ}$, 157.5$^{\circ}$\} (e.g. by applying an offset
angle to the polarization compensation controlled via the
reference laser beam). The data set of 16 different polarization
correlations between the two ground stations already allows a
test of the violation of a Bell inequality of the CHSH
type~\cite{CHSH69}. In addition, strict Einstein locality 
conditions can be obeyed by randomly switching the 
measurement basis at both receiver stations~\cite{Weihs98a}.

The second stage of the experiment is a combination of the
measurements described above. With equivalent orientation of the
analyzer modules in the \{0$\,^{\circ}$,90$\,^{\circ}$\} and
\{45$\,^{\circ}$/135$\,^{\circ}$\} bases at both receiver
stations, the coincidence events measured in the same basis
(e.g.~in the \{45$\,^{\circ}$,135$\,^{\circ}$\} basis for both
receiver stations) allow to establish a quantum cryptographic key
analogous to Experiment~1 (see
Section~\ref{subsec:description_exp1}). Additionally, both of the
receiver stations are allowed to randomly rotate their analyzer
basis by 22.5$\,^{\circ}$, thus also performing a test of Bell's
inequality. The violation of a Bell inequality is sufficient as a
security proof of the quantum key distribution
protocol~\cite{Fuchs97a,Scarani01a,Acin03}. Net key bit rate, QBER
and degree of security of the quantum key distribution will be
evaluated. Given the above approximations, a raw key generation
rate of 600~bits per link duration of approx.~5 minutes might be
expected.

\section{Conclusions}
We conclude that present day technology enables us to bring
quantum entanglement into space thus taking advantage of this
unique "lab" environment. It allows us to perform fundamental
tests of quantum physics, above all a test of Bell's inequality,
at distances far beyond the capabilities of Earth-bound
laboratories. In the first stages, those distances might not be
astronomical, as would be the case when using a flotilla of
satellites, but already with the use of a LEO-based transmitter
and two Earth-bound receivers one can overcome the Earth-bound
limitations by several orders of magnitude. Although there exists
not yet a space-qualified source of entangled photons, the system
complexity of present diode setups is sufficiently low to
consider space qualification a feasible task. In the long run, 
placing both the source and the receivers in space additionally 
opens up the possibilities to perform novel tests of quantum physics making
specific use of the added value of space.

\acknowledgments This paper evolved from a project supported by
the European Space Agency under ESTEC/Contract No.
16358/02/NL/SFe, "Quantum Communications in Space". We wish to
thank Josep Maria Perdigues Armengol for monitoring and
supporting this work. We further acknowledge support by the
Austrian Science Foundation (FWF) and the Alexander von Humboldt
Foundation.

\bibliography{qspace,aspelmeyer,qspace_spie}
   \bibliographystyle{ieeetr}

\end{document}